\documentclass[aps,prl,twocolumn,superscriptaddress]{revtex4}
\usepackage{epsfig}
\usepackage{graphics}
\begin{document}
\title{Isospin Dependence in the Odd-Even Staggering of Nuclear Binding Energies\footnote{Physical Review Letters 95 (2005) 042501, http://prl.aps.org/}}
\author{Yu.A.~Litvinov}%
\affiliation{Gesellschaft f\"ur Schwerionenforschung GSI,
Planckstra{\ss}e 1, 64291
Darmstadt, Germany}%
\affiliation{II. Physikalisches Institut, JLU Giessen,
Heinrich-Buff-Ring 16, 35392 Giessen, Germany}%
\author{T.J.~B\"urvenich} \affiliation{Los Alamos National Laboratory, Los
Alamos, NM 87545, USA} \affiliation{Max-Planck-Institut f\"ur
Kernphysik, Saupfercheckweg 1, 69117 Heidelberg, Germany}
\author{H.~Geissel}
\affiliation{Gesellschaft f\"ur Schwerionenforschung GSI,
Planckstra{\ss}e 1, 64291 Darmstadt, Germany} \affiliation{II.
Physikalisches Institut, JLU Giessen, Heinrich-Buff-Ring 16, 35392
Giessen, Germany}
\author{Yu.N.~Novikov}
\affiliation{St.~Petersburg Nuclear Physics Institute, 188350
Gatchina, Russia}
\author{Z.~Patyk}
\affiliation{Soltan Institute for Nuclear Studies, Hoza 69, 00681
Warsaw, Poland}
\author{C.~Scheidenberger}
\author{F.~Attallah}%
\affiliation{Gesellschaft f\"ur Schwerionenforschung GSI,
Planckstra{\ss}e 1, 64291 Darmstadt, Germany}
\author{G.~Audi}%
\affiliation{CSNSM-IN2P3-CNRS, B\^atiment 108, 91405 Orsay Campus, France}%
\author{K.~Beckert}%
\author{F.~Bosch}%
\affiliation{Gesellschaft f\"ur Schwerionenforschung GSI,
Planckstra{\ss}e 1, 64291 Darmstadt, Germany}
\author{M.~Falch} \affiliation{Sektion Physik, LMU M\"unchen, Am Coulombwall,
85748 Garching, Germany}
\author{B.~Franzke}
\affiliation{Gesellschaft f\"ur Schwerionenforschung GSI,
Planckstra{\ss}e 1, 64291 Darmstadt, Germany}
\author{M.~Hausmann}
\affiliation{Los Alamos National Laboratory, Los Alamos, NM 87545,
USA}
\author{Th.~Kerscher}
\affiliation{Sektion Physik, LMU M\"unchen, Am Coulombwall, 85748
Garching, Germany}
\author{O.~Klepper}
\author{H.-J.~Kluge}
\author{C.~Kozhuharov}
\affiliation{Gesellschaft f\"ur Schwerionenforschung GSI,
Planckstra{\ss}e 1, 64291 Darmstadt, Germany}
\author{K.E.G.~L\"obner}
\affiliation{Sektion Physik, LMU M\"unchen, Am Coulombwall, 85748
Garching, Germany}
\author{D.G.~Madland}
\affiliation{Los Alamos National Laboratory, Los Alamos, NM 87545, USA}
\author{J.A.~Maruhn}
\affiliation{Institut f{\"u}r Theor. Physik, JWGU Frankfurt,
Robert-Mayer-Stra{\ss}e 10, 60054 Frankfurt, Germany}
\author{G.~M{\"u}nzenberg}
\affiliation{Gesellschaft f\"ur Schwerionenforschung GSI,
Planckstra{\ss}e 1, 64291 Darmstadt, Germany}
\author{F.~Nolden}
\affiliation{Gesellschaft f\"ur Schwerionenforschung GSI,
Planckstra{\ss}e 1, 64291 Darmstadt, Germany}
\author{T.~Radon}
\author{M.~Steck}
\author{S.~Typel}
\affiliation{Gesellschaft f\"ur Schwerionenforschung GSI,
Planckstra{\ss}e 1, 64291 Darmstadt, Germany}
\author{H.~Wollnik}\affiliation{II. Physikalisches Institut, JLU Giessen, Heinrich-Buff-Ring 16, 35392 Giessen, Germany}
\begin{abstract}
The FRS-ESR facility at GSI provides unique conditions for
precision measurements of large areas on the nuclear mass surface
in a single experiment. Values for masses of 604 neutron-deficient
nuclides (30$\le$Z$\le$92) were obtained with a typical
uncertainty of 30~$\mu{u}$. The masses of 114 nuclides were
determined for the first time. The odd-even staggering (OES) of
nuclear masses was systematically investigated for isotopic chains
between the proton shell closures at Z=50 and Z=82. The results
were compared with predictions of modern nuclear models. The
comparison revealed that the measured trend of OES is not
reproduced by the theories fitted to masses only. The spectral
pairing gaps extracted from models adjusted to both masses and
density related observables of nuclei agree better with the
experimental data.
\end{abstract}
\pacs{}\keywords{}
\maketitle
\par%
Significant progress has been achieved over the last years in
constructing self-consistent mass
models~\cite{Bender_RMP,Lunney_RMP}. These models aim to reliably
describe the properties of nuclei far off the valley of
$\beta$-stability, where the experimental information is scarce or
even not available yet. For instance, in modelling the
astrophysical r-process of nuclear synthesis one needs precise
knowledge of masses and half-lives of very exotic nuclei and one
has to rely on theoretical predictions since most of the nuclides
involved have not even been produced in the laboratory yet. The
predictions for these nuclides dramatically deviate for the
different models~\cite{Lunney_RMP}. Thus new experimental data on
exotic nuclei and consequently better understanding of nuclear
structure away from the valley of $\beta$-stability is essential
for further theoretical development.
\par%
Odd-even staggering of nuclear binding energies (OES) was detected
in the early days of nuclear physics~\cite{He-ZP78} and was
explained by the presence of pairing correlations between nucleons
in the nucleus~\cite{Bohr}. Pairing contributes only little to the
total nuclear binding energy but its influence on the nuclear
structure is significant.
\par%
The common way to extract experimental information about the
pairing correlations is to measure the value of the OES which
approximates the pairing-gap energy ($\Delta$) in the standard
Bardeen-Cooper-Schrieffer (BCS) theory~\cite{Bardeen-PR108}. The
latter quantity is connected with the strength of the pairing
interaction ($G$):
\begin{equation}\label{bcs_gap1}
  \frac{2}{G} = \sum_{\nu}
  \frac{1}{\sqrt{\left(\varepsilon_{\nu}-\lambda\right)^2+\Delta^2}},
\end{equation}
where $\varepsilon_{\nu}$ is the single-particle energy and
$\lambda$ is the chemical potential. The summation goes over all
single-particle levels $\nu$ below and above the Fermi energy. In
order to evaluate this sum in local pairing functionals a (smooth)
cut-off in energy is usually implemented.
\par%
Neutron ($\Delta_n$) and proton ($\Delta_p$) pairing gaps are
usually determined from finite-difference equations of measured
masses \cite{Madland-Nix}, e.g. by the five-point formulae:
\begin{eqnarray}
\nonumber\Delta^{(5)}_n=
-\frac{1}{8}~[M(Z,N+2)-4M(Z,N+1)+\\
6M(Z,N)-4M(Z,N-1)+M(Z,N-2)], \label{delta5n}
\end{eqnarray}
\begin{eqnarray}
\nonumber\Delta^{(5)}_p=
-\frac{1}{8}~[M(Z+2,N)-4M(Z+1,N)+\\
6M(Z,N)-4M(Z-1,N)+M(Z-2,N)], \label{delta5p}
\end{eqnarray}
where M(Z,N) is the mass of an atom with Z protons and N neutrons.
\par%
The well-known parametrization
$\Delta{\simeq}12/\sqrt{A}$~MeV~\cite{Bohr} ($A=N+Z$) provides the
average trend for nuclei close to stability. A dependence of the
pairing strength on the neutron excess was suggested in
\cite{Nilsson-NPA131}. It was later observed from the mass
determination of exotic Dy-Hg isotopes that $\Delta_p$ and
possibly $\Delta_n$ increase towards the proton drip-line
\cite{Alkhazov-ZP311}.
\par%
In this letter we present new results on the OES obtained from our
high-precision mass measurements compared with predictions of
modern nuclear theories.
\par%
The experiment for direct mass measurements was performed at the
FRS-ESR facility as continuation of a successful scientific
program addressing basic nuclear properties of stored exotic
nuclides~\cite{Raidi-PRL,Raidi,Novikov}. Exotic nuclei were
produced by projectile fragmentation of a (600-900)~MeV/u
$^{209}$Bi primary beam in (4-8)~g/cm$^2$ beryllium targets placed
at the entrance of the fragment separator (FRS)~\cite{Ge-NIM24}.
The fragments were spatially separated in-flight and injected into
the cooler-storage ring ESR~\cite{Franzke}. In the ESR, the
velocity spread of the stored fragments was reduced by electron
cooling to $\delta{v}/v\approx5\cdot{10}^{-7}$. This condition
provides an unambiguous relation between the revolution
frequencies of the ions and their mass-to-charge ratios which is
the basis for Schottky Mass Spectrometry (SMS). The time required
for the electron cooling was about 10~s, which constrains the
range of nuclides that might be investigated by this method. The
SMS has reached the ultimate sensitivity by recording single ions
stored in the ESR leading to a mass resolving power of more than
2$\cdot$10$^6$ (FWHM) \cite{Li-NN,HG-RNB6}.
\par%
In this new experiment with SMS, 582 different nuclides were
observed in the frequency spectra. From this set of nuclei, 117
were used for calibration. The achieved mass accuracy was
typically 30~$\mu{u}$ which represents an improvement by a factor
of three compared to our former experiments~\cite{Raidi}. In
addition, the masses of 139 nuclides were determined indirectly by
means of known decay energies ($\alpha$, $\beta$, or proton
emission). The masses of 114 nuclides were obtained for the first
time~\cite{Li-NN}. The measured masses cover a large area of
neutron-deficient nuclides from krypton to uranium. All {\it
directly measured} values (see \cite{Li-97}) have been included in
the latest Atomic Mass Evaluation~\cite{AW03}.
\par%
The achieved experimental mass accuracy has allowed us to perform
new systematical studies on nuclear pairing. The new data were
combined with the data of Ref.~\cite{AW} and precise values of OES
for {\it all} even-Z isotopic chains in the region between the
Z=50 and Z=82 closed shells were extracted.
Only even-even nuclei were considered. The results obtained show
that indeed the values of OES for both the protons and the
neutrons increase towards the proton drip-line, thus confirming
earlier observations for a small number of nuclides
\cite{Alkhazov-ZP311}. {\it Moreover, this is a general trend for
all even-Z isotopic chains from tin to lead.} The tin, tellurium,
mercury, and lead isotopes were not considered for protons since
the closed shells at Z=50 and Z=82 have strong influence.
Similarly, the nuclides with N=80, 82, 84, 124, 126, and 128 were
excluded for neutrons. It is necessary to note that no such
general trend of the OES was observed for isotopes below tin. For
isotopes above lead the experimental information is still too
scarce to draw a definite conclusion. The large number of newly
obtained OES values allow us to perform quantitative comparisons
with calculations. We will first compare the results with a
macroscopic-microscopic model and then continue with up-to-date
microscopic models.
\par%
\begin{figure}[t!]
\includegraphics*[width=8.0cm]{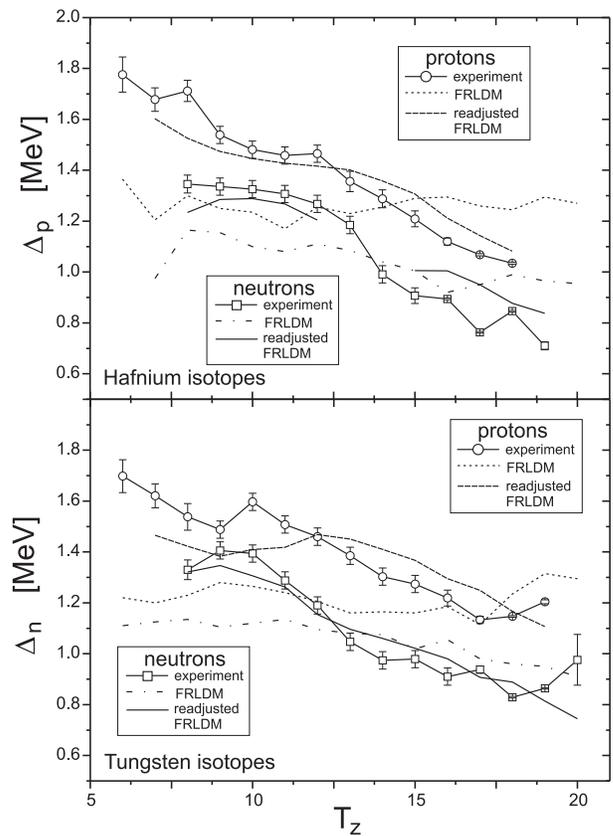}
\caption {Comparison of the proton and neutron pairing-gap
energies for even-even hafnium (upper panel) and tungsten (lower
panel) isotopes derived with 2nd-order mass differences of
experimental masses and from the predictions of the original FRLDM
mass model \cite{Mo-ADNDT59} and with newly readjusted pairing
strengths. The experimental values are taken from
Refs.~\cite{Li-97,AW}.} \label{frldm}
\end{figure}
\begin{figure}[t!]
\begin{center}
\includegraphics*[width=8.0cm]{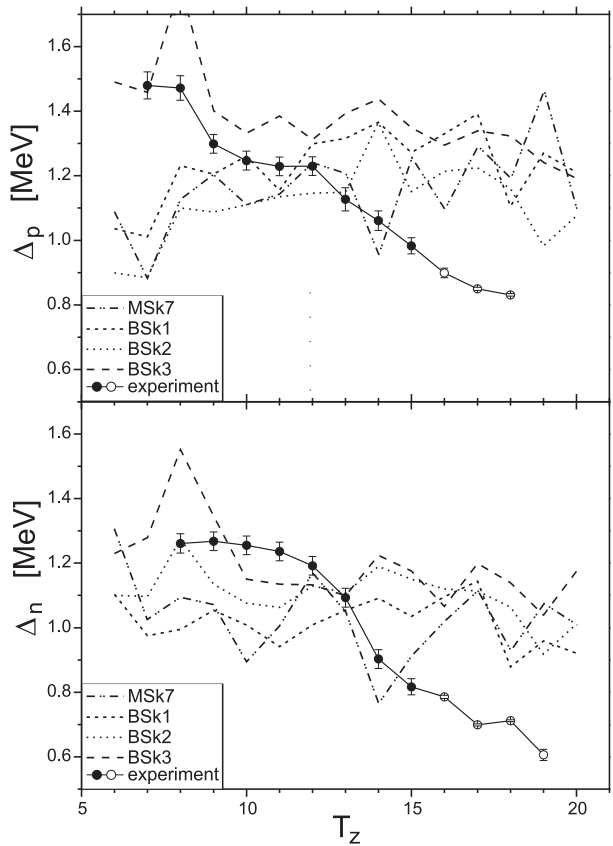}
\caption {Comparison of the proton ($\Delta_p$) and neutron
($\Delta_n$) pairing-gap energies for even-even hafnium isotopes
derived from fourth-order differences of experimental and of
calculated masses. Hartree-Fock plus pairing treated with the BCS
formalism (MSk7) and three HF-Bogoliubov models with different
cut-off parameterizations (BSk1 and BSk2) and density dependent
pairing (BSk3) were used. The experimental values are taken from
this work (full symbols) and from Ref.~\cite{AW} otherwise.}
\label{pair}
\end{center}
\end{figure}
\begin{figure}[t!]
\begin{center}
\includegraphics*[width=8.0cm]{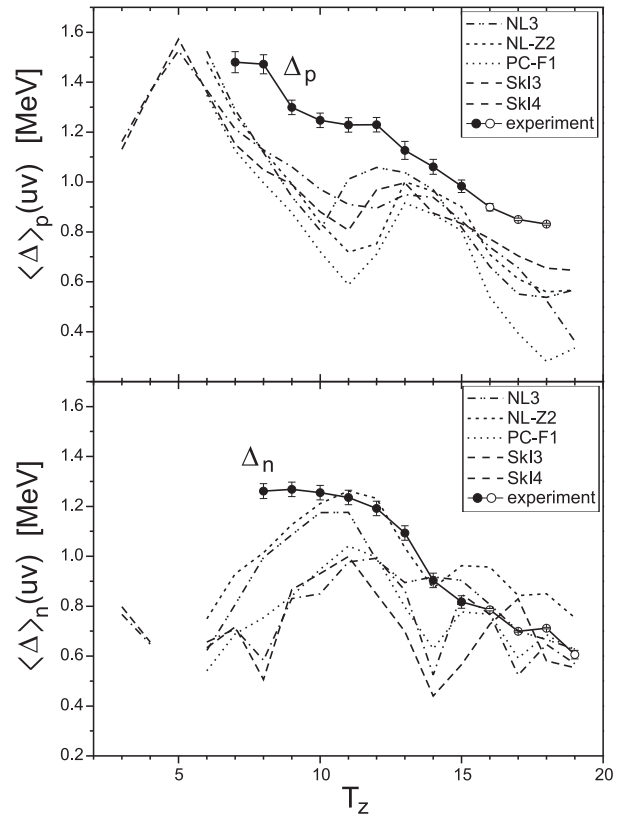}
\caption {Comparison of the proton ($\langle\Delta\rangle_p (uv)$)
and neutron ($\langle\Delta\rangle_n (uv)$) pairing-gap energies
for even-even hafnium nuclides calculated with several RMF and
Skyrme-Hartree-Fock models (see text). These models were adjusted
not only to the nuclear binding energies but also to form-factor
related observables. The experimental points were derived from
measured masses using Eqs. (\ref{delta5n},\ref{delta5p}). The full
symbols represent our new experimental values, the others are from
Ref.~\cite{AW}.} \label{rmf}
\end{center}
\end{figure}
The pairing-gap energies were extracted from the masses calculated
with the Finite-Range Liquid-Drop Model (FRLDM) \cite{Mo-ADNDT59}.
In this model, the single-particle potential generated by the
Yukawa interaction was used for the microscopic part. To obtain a
maximal number of extracted OES values,
the calculations were done using 2nd-order mass differences
(three-point formulae \cite{Madland-Nix}). The comparison with the
experiment is shown for the isotopic chains of hafnium and
tungsten in Fig. \ref{frldm}. It is clearly seen that the
experimental isospin dependence of pairing-gap energies is not
reproduced by the original FRLDM \cite{Mo-ADNDT59}. To improve
this description, the BCS pairing part of the model has been
adjusted to the new experimental data. Different from Ref.
\cite{Mo-ADNDT59}, a single-particle spectrum was generated with
the deformed Woods-Saxon potential. The pairing strength $G$ was
parameterized with 2 constants for protons ($p$) and neutrons
($n$) $G_{p(n)}=g_{0p(n)}/A+g_{1p(n)}(N-Z)/A^2$
\cite{Nilsson-NPA131,Madland-Nix}. The number of levels taken into
account (see Eq. \ref{bcs_gap1}) was equal to N for neutrons and Z
for protons. All experimental OES values for even-even nuclides
between Z=50 and Z=82 were used for the adjustment. The best
agreement between the values of OES derived from the calculated
masses and experimental values was achieved with
$g_{0p}=g_{0n}=20.80$~MeV and
$g_{1p}=-g_{1n}=22.40$~MeV~\cite{Li-NN}. Note that the constants
($g_{0p}$, $g_{0n}$) and ($g_{1p}$, $g_{1n}$) converge to the same
values, which is quite remarkable since this was not a constraint
demanded in the analysis. The results obtained are labelled in
Fig. \ref{frldm} as {\it readjusted FRLDM}. The difference between
the original FRLDM of Ref.~\cite{Mo-ADNDT59} and the readjusted
model is obvious. With this new pairing description the
$\sigma_{rms}$ value for the prediction of nuclear binding
energies between Z=50 and Z=82 closed shells has improved by about
25~\%.
\par%
Going to microscopic models, pairing-gap energies were calculated
from predictions of nuclear mass calculations of several
self-consistent mass models, which are Skyrme-Hartree-Fock (SHF)
calculations plus pairing treated in the BCS (HF+BCS) formalism
(MSk7 force) \cite{HFBCS1} and three models where pairing is
treated with the Bogoliubov (HFB) approach with different pairing
cut-off parameterizations (BSk1, BSk2) \cite{{HFB1},{HFB2}} and
including density-dependent pairing (BSk3)~\cite{HFB3}. The
results calculated with 4th-order mass differences
(Eqs.~\ref{delta5n},\ref{delta5p}) are presented for even hafnium
isotopes in Fig.~\ref{pair}. It is clearly seen that the general
trend is {\it not} reproduced by any of the models. A comparison
with four other recent HFB models where the implementation of
different effective masses, with and without density-dependent
pairing (BSk4-BSk7)~\cite{HFB4}, showed similar results as those
of Fig.~\ref{pair}.
\par%
Another branch of self-consisistent mean-field models are
relativistic mean-field (RMF) models employing finite-range meson
fields (FR) or point couplings (PC). For this study, we employed
three of the best parameterizations available, namely
NL-Z2~\cite{Th-1} and NL3~\cite{Th-2} for the finite range
variant, and PC-F1~\cite{Th-3} for the point-coupling model.
In contrast to the mass models described above, these RMF forces
are adjusted to both energy and form-factor related observables
(e.g. rms radii, diffraction radii, surface thicknesses, etc.),
and are meant to describe both kinds of observables. Furthermore,
we performed calculations with the SHF forces SkI3 and SkI4
\cite{SkI} with extended spin-orbit terms, which -- similar to the
RMF forces -- have been adjusted to both masses and density
related observables. In both the RMF and the SHF models, we employ
BCS pairing with a density-independent $\delta$-force. As a first
approach, the pairing gaps have been estimated from the
single-particle spectrum with {\it uv}-weighted single-particle
gaps \cite{Bender_EPJ8} ($v^2$ are the occupation probabilities),
which circumvents the uncertainties related to the calculation of
odd-even systems. These quantities constitute a measure of the
pairing contribution to the OES. As discussed in Ref.
\cite{Bender_EPJ8}, these results need to be carefully interpreted
due to polarization effects and the non-pairing-type contributions
to the OES.
\par%
It is striking that the RMF and SHF calculations in Fig.~\ref{rmf}
give very similar results: the general trend of the rising
pairing-gap energies is reproduced. However, some local
discrepancies are observed, as e.g. close to $T_z$=5, which can be
related to the N=82 closed shell. Although the models in
Fig.~\ref{pair} in general have much higher predictive power for
the nuclear masses \cite{Lunney_RMP} the difference in the
description of the experimental OES data is obvious. This result
has not been expected and demonstrates the need for a better
understanding of both the roles of the various observables and
adjustment protocols as well as the procedure of calculating OES
within these frameworks.
\par%
All models tested in this letter take into account the nuclear
deformation which is essential here since most of the nuclei
investigated are deformed. Moreover, the observed general trend of
OES has the same slope and magnitude for nearly all isotopic
chains from xenon to platinum. Use of the 2nd, 3rd, or 4th-order
mass differences to disentangle the mean field contributions to
the OES and pairing-gap energies has been intensively discussed in
the literature~\cite{{Bender_EPJ14},{DobaczewskiPRC63},
{Duguet_PRC65},{RutzPLB},{SatulaPRL}}. Our overall conclusions, as
checked in different analyses, are not changed if 3, 4, or 5-mass
formulae are used. Since the mass number changes within an
isotopic chain, volume effects might contribute to the observed
isospin dependence of the OES.
\par%
With the new data available, it became possible to examine the OES
predictions of different theories. The new results are helpful for
a better description of the pairing in exotic nuclei, which is
mandatory for a {\it reliable} theory.
\par%
An important future aspect is whether the observed trend of the
pairing-gap energies persists for nuclides with even greater
neutron excess. A recent experiment has been performed at the
FRS-ESR to measure masses in the Yb-Pb region on the neutron-rich
side of the chart of nuclides. Measurements of very exotic
neutron-rich nuclides, which cannot be produced with the present
facility, are foreseen within the FAIR project \cite{CDR}.

\section{Acknowledgments}
The authors thank P.--G. Reinhard and R. Wyss for helpful
discussions, M. Samyn for providing the HFB calculations, as well
as M. Bender for help with coding problems.


\begin{thebibliography}{99}
\bibitem{Bender_RMP}
M.~Bender et al., Rev. Mod. Phys. 75 (2003) 121.
\bibitem{Lunney_RMP}
D.~Lunney et al., Rev. Mod. Phys. 75 (2003) 1021.
\bibitem{He-ZP78}
W.~Heisenberg, Z. Phys. 78, 156 (1932).
\bibitem{Bohr}
A.~Bohr, B.R.~Mottelson, {\it Nuclear Structure}, World Scientific
Publ., Singapore, 1998.
\bibitem{Bardeen-PR108}
J.~Bardeen et al., Phys. Rev. 108, 1175 (1957).
\bibitem{Madland-Nix}
D.G.~Madland, J.R.~Nix, Nucl. Phys. A476 (1988) 1.
\bibitem{Nilsson-NPA131}
S.G.~Nilsson et al., Nucl. Phys. A131 (1969) 1.
\bibitem{Alkhazov-ZP311}
G.D.~Alkhazov et al., Z. Phys. A311 (1983) 245.
\bibitem{Raidi-PRL}
T.~Radon et al., Phys. Rev. Lett. 78 (1997) 4701.
\bibitem{Raidi}
T.~Radon et al., Nucl. Phys. A677 (2000) 75.
\bibitem{Novikov}
Yu.N. Novikov et al., Nucl. Phys. A697 (2002) 92.
\bibitem{Ge-NIM24}
H.~Geissel et al., Nucl. Instr. and Meth. B70 (1992) 286.
\bibitem{Franzke}
B.~Franzke, Nucl. Instr. and Meth. B24/25 (1987) 18.
\bibitem{Li-NN}
Yu.A.~Litvinov et al., Nucl. Phys. A734 (2004) 473.
\bibitem{HG-RNB6}
H.~Geissel et al., Nucl. Phys. A746 (2004) 150c.
\bibitem{Li-97}
Yu.A.~Litvinov et al., Nucl. Phys. A756 (2005) 3.
\bibitem{AW03}
G.~Audi et al., Nucl. Phys. A729 (2003) 3.
\bibitem{AW}
G.~Audi et al., Nucl. Phys. A624 (1997) 1.
\bibitem{Mo-ADNDT59}
P.~M{\"o}ller et al., ADNDT 59 (1995) 185.
\bibitem{HFBCS1}
S.~Goriely et al., ADNDT 77 (2001) 311.
\bibitem{HFB1}
M.~Samyn et al., Nucl. Phys. A700 (2002) 142.
\bibitem{HFB2}
S.~Goriely et al., Phys. Rev. C66 (2002) 024326.
\bibitem{HFB3}
M.~Samyn et al., Nucl. Phys. A725 (2003) 69.
\bibitem{HFB4}
S.~Goriely et al., Phys. Rev. C68 (2003) 054325.
\bibitem{Th-1}
M. Bender et al., Phys. Rev. C 60 (1999) 34304
\bibitem{Th-2}
G.A. Lalazissis et al., Phys. Rev. C 55 (1997) 540
\bibitem{Th-3}
T. B{\"u}rvenich et al., Phys. Rev. C 65, 044308 (2002)
\bibitem{SkI}
P.-G. Reinhard, H. Flocard, Nucl Phys. {A584} (1995) 467
\bibitem{Bender_EPJ8}
M.~Bender et al., Eur. Phys. J. A8 (2000) 59.
\bibitem{Bender_EPJ14}
M.~Bender et al., Eur. Phys. J. A 14 (2002) 23.
\bibitem{Duguet_PRC65}
T.~Duguet et al., Phys. Rev. C65 (2001) 014311.
\bibitem{SatulaPRL}
W.~Satula et al., Phys. Rev. Lett. 81 (1998) 3599.
\bibitem{RutzPLB}
K.~Rutz et al. Phys. Lett. B468 (1999) 1.
\bibitem{DobaczewskiPRC63}
J.~Dobaczewski et al., Phys. Rev. C63 (2001) 024308.
%
\bibitem{CDR} W. Henning, Nucl. Phys. A734 (2004)
654.
%
\end{thebibliography}
\end{document}